\documentclass{jfm}

\usepackage{graphicx}
\usepackage{newtxtext}
\usepackage{newtxmath}
\usepackage{natbib}
\usepackage{hyperref}
\hypersetup{
    colorlinks = true,
    urlcolor   = blue,
    citecolor  = black,
}

\newcommand{\RomanNumeralCaps}[1]

\definecolor{tomato}{RGB}{9,121,105}

\title{Viscoelastic wetting: Cox-Voinov theory with normal stress effects}

\author{Minkush Kansal\aff{1}
  \corresp{\email{m.kansal@utwente.nl}},
  Vincent Bertin\aff{1},
  Charu Datt\aff{1, 2},
  Jens Eggers \aff{3}
 \and Jacco H. Snoeijer\aff{1}}

\affiliation{\aff{1}Physics of Fluids Group, Faculty of Science and Technology, University of Twente, P.O. Box 217, 7500 AE Enschede, The Netherlands
\aff{2}Max Planck Institute for the Physics of Complex Systems, N\"{o}thnitzer Str. 38, 01187 Dresden, Germany
\aff{3}School of Mathematics, University of Bristol, Fry Building, Woodland Road, Bristol BS8 1UG, UK}

\begin{document}
\maketitle

\begin{abstract}
The classical Cox-Voinov theory of contact line motion provides a relation between the macroscopically observable contact angle, and the microscopic wetting angle as a function of contact line velocity. Here we investigate how viscoelasticity, specifically the normal stress effect, modifies wetting dynamics. Using the thin film equation for the second-order fluid, it is found that the normal stress effect is dominant at small scales \textcolor{black}{yet can significantly affect macroscopic motion}. We show that the effect can be incorporated in the Cox-Voinov theory through an apparent microscopic angle, which differs from the true microscopic angle. The theory is applied to the classical problems of drop spreading and dip-coating, which shows how normal stress facilitates (inhibits) the motion of advancing (receding) contact lines. For rapid advancing motion, the apparent microscopic angle can tend to zero, in which case the dynamics is described by a regime that was already anticipated in~\citet{boudaoud2007non}.
\end{abstract}




\section{Introduction}
\label{sec:introduction}

\begin{figure}
	\centerline{\includegraphics[width=\linewidth]{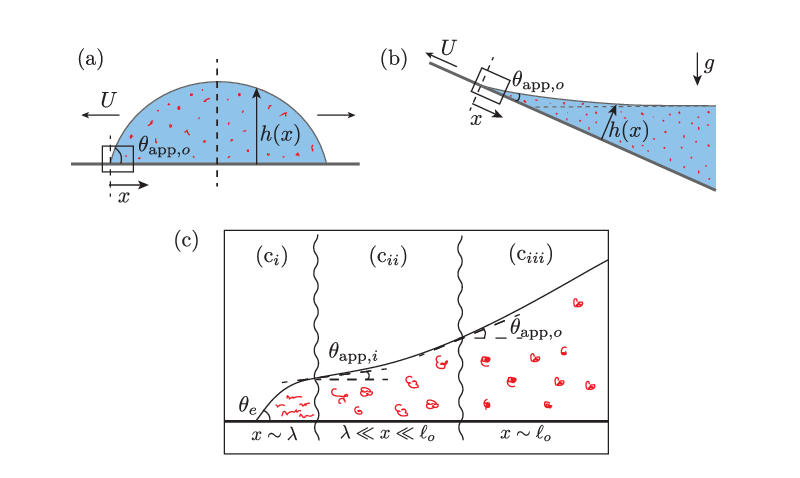}}
	\caption{Moving contact lines for viscoelastic liquids encountered in drop spreading (a) and dip-coating (b). At \textcolor{black}{a} large scale, the flow is characterised by a macroscopic apparent contact angle $\theta_{{\rm app},o}$, seen at an outer scale $\ell_o$. (c) Microscopic view of the interface. $(\mathrm{c}_i)$ the highest shear rates are encountered close to the contact line (slip length scale $\lambda$), where polymers become highly stretched. This gives rise to large normal stresses that ``bend" the interface to angles below the equilibrium contact angle $\theta_e$. In this paper we introduce $\theta_{{\rm app},i}$ as the resulting apparent microscopic angle. $(\mathrm{c}_{ii})$ On a scale-free region $\lambda \ll x \ll \ell_o$, the interface slope follows the Voinov solution reflecting the visco-capillary balance; $\theta_{{\rm app},i}$ serves as the apparent inner boundary condition at small scale. $(\mathrm{c}_{iii})$ At large scales, the slope is given by the apparent outer angle $\theta_{\mathrm{app},o}$. 
		}
	\label{fig:fig1}
\end{figure}

The motion of a contact line governs a variety of phenomena, such as drops moving on surfaces, drop spreading, and coating applications. This problem has been \textcolor{black}{widely} explored for Newtonian fluids over the last decades~\citep{gennes2004capillarity,bonn2009wetting,snoeijer2013moving}. Within the hydrodynamic framework of wetting, the macroscopic flow is often described using the Cox-Voinov theory \citep{cox_1986, voinov1976hydrodynamics}, which connects the macroscopic contact angle to the microscopic angle. The relation between these angles is intricate, since the balance of the viscous shear forces (viscosity $\eta$) and capillary forces (surface tension $\gamma$) leads to a ``bending'' of the interface in a region near the contact line. Macroscopically, the result can be expressed as a relation for the apparent outer (macroscopic) angle $\theta_{\mathrm{app},o}$, defined in Figure~\ref{fig:fig1}(a,b) for droplet spreading and dip-coating, respectively. An explicit treatment of the problem is often given using the thin film (lubrication) framework, see for e.g., \cite{eggers2015singularities}, which leads to a dependence of the macroscopic angle on the contact line speed $U$ through the form \citep{cox_1986, voinov1976hydrodynamics}:

\begin{equation}
	\label{eq:Ca_and_N_definition}
\theta_{\mathrm{app},o}^3 = \theta_e^3 \pm 9\mathrm{Ca} \ln\left(\ell_o/\ell_i\right), \quad \quad \mathrm{Ca} = \frac{\eta U}{\gamma}.
\end{equation}
Here $\theta_e$ is the (microscopic) equilibrium angle, and $\mathrm{Ca}$ the capillary number. The ``$\pm$" depends on whether the contact line is advancing (+) or receding (-). The formula contains an inner length scale $\ell_i$, necessary to regularise the moving contact line singularity \citep{huh1971hydrodynamic}, as well as an outer length scale $\ell_o$ that reflects the macroscopic geometry of the problem (\textit{e.g.}, the radius of a spreading drop). The strict validity of \eqref{eq:Ca_and_N_definition} is restricted to small contact angles and small capillary numbers, due to assumptions underlying the analysis \citep{cox_1986, voinov1976hydrodynamics, eggers2015singularities}. Yet, it has been successfully applied to a large class of problems \citep{snoeijer2013moving}.

Many applications like the deposition of pesticides on plant leaves or ink-jet printing involve polymer solutions. Such fluids exhibit non-Newtonian rheology such as shear-thinning viscosity and viscoelasticity, which can affect the contact line dynamics as has been demonstrated through several experimental studies  \citep{wei2009a, wei2009dynamic, wang2007spreading, seevaratnam2007, han2013spreading, Kim2015Mass}. For the canonical problem of liquid spreading on a substrate, it was found that a non-Newtonian drop \textcolor{black}{spreading} on completely wetting surfaces still followed the long-time power-law dynamics $R \sim t^{n}$, where $R(t)$ is the drop radius, and the exponent $n$ is slightly smaller than the Cox-Voinov-Tanner's law $n=1/10$~\citep{rafai2004spreading}. This comparatively weak effect could be attributed to both shear-thinning viscosity and the consequence of normal stress effects. During the very early stages of drop spreading, ensuing after a droplet is gently brought into contact with a \textcolor{black}{substrate}, the spreading exhibits much larger exponents, with $n=1/2$ for low-viscosity Newtonian drops \citep{Biance2007}. Interestingly, the addition of polymers has no measurable effect on this rapid  spreading~\citep{bouillant2022rapid,yada2023rapid}. Similarly, during drop impact, the initial spreading phase is nearly unaffected by the presence of polymers~\citep{bartolo2007dynamics,gorin2022universal,sen2022elastocapillary}. 

By contrast, the retraction of a droplet, \textit{e.g.} after an impact, is strongly influenced by the presence of polymers~\citep{bergeron2000controlling}. The typical retraction velocity was found to decrease with the (first) normal stress coefficient $\psi$, following a scaling $U \sim 1/\sqrt{\psi}$~\citep{bartolo2007dynamics}. This result can be interpreted from the balance between the normal stress near the contact line $\sim \psi \left(U/\ell \right)^2$ and the Laplace pressure $\gamma h'' \sim \gamma \theta_e^2/\ell$, giving a typical velocity

\begin{equation}\label{eq:upsi}
U_\psi = \theta_e \sqrt{\frac{\gamma \ell}{\psi}},
\end{equation}
where $\ell$ is to \textcolor{black}{be} interpreted as a microscopic cutoff length~\citep{bartolo2007dynamics}, very much like $\ell_i$ in \eqref{eq:Ca_and_N_definition}. The best-fit of experimental data gave $\ell$ to be of the order of tens of microns~\citep{bartolo2007dynamics}, which is larger than the expected nanometric length used in Newtonian fluids. \textcolor{black}{Alternatively, this result can be obtained from a balance of capillary force $\gamma \left( \cos{\theta} - \cos{\theta_e} \right) \sim \gamma \theta_e^2/2$, and viscoelastic force $\psi U^2/ \ell$ induced by the extra tension along the streamlines~\citep{bartolo2007dynamics}.} The addition of polymers also changes the drop shedding behind a sliding droplet \citep{xu2018viscoelastic}.

\textcolor{black}{In fact,} polymers have been observed to be strongly stretched at \textcolor{black}{a} contact line (during retraction) \citep{smith2010effect} and stretched and depleted at the contact line (during spreading) \citep{Shin2016} and are sometimes left on the surface after the retraction~\citep{zang2013impact}. This could be induced by either non-hydrodynamic polymer-surface interactions \citep{guyard2021near}, or shear-induced migration \citep{han2013spreading, Shin2015}. However, it has remained a challenge to disentangle the possible roles of all these mechanisms, which in part is due to the lack of a hydrodynamic framework for viscoelastic wetting. 

The aim of the present study is to \textcolor{black}{explore systematically} the influence of viscoelastic normal stress on the motion of contact lines. There have been several studies dedicated to quantifying normal stress effects in wetting flows. Numerical schemes have been developed using the phase-field method and polymer rheological models like Oldroyd-B or Giesekus \citep{yue2012, wang2015dynamic,wang2017impact} and results using these were found to be consistent with prior experimental observations \citep{wei2009dynamic, han2013spreading}, in that the viscoelastic stresses, confined to a small region near the contact line, had only a mild effect on the contact-line motion of spreading drops. \citet{boudaoud2007non} introduced a phenomenological thin-film model that includes normal stress effects. Solving the resulting equation in the specific case of complete wetting, the advancing contact angle is found to follow 
\begin{equation}
\label{eq:Boudaoud}
	\theta_{\mathrm{app},o}^3 = 9 \mathrm{Ca} \ln{\left( \frac{\ell_o}{\ell_\mathrm{VE}}\right)}, \quad \quad \ell_\mathrm{VE} = \frac{\psi U}{2 \eta}.
\end{equation}
This expression can be seen as a form of the Cox-Voinov law \eqref{eq:Ca_and_N_definition}, with $\theta_e=0$ and the microscopic regularizing length provided by a viscoelastic length $\ell_\mathrm{VE}$. The latter corresponds to the product between the contact line velocity $U$ and the typical relaxation time of the liquid $\psi/\eta$. \textcolor{black}{However, it has remained unclear whether this theory can actually be derived from any tensorial constitutive relation, and how these results carry over to partial wetting conditions, for which the equilibrium contact angle $\theta_e$ must play a role. Also, as the equation derived by~\citet{boudaoud2007non} (given as equation~\eqref{eq:Boudaoud}) does not admit any solutions with partial wetting conditions, it cannot be applied to receding contact lines - which is crucial to understand the observed strong effects of normal stresses on the retraction velocity~\citep{bartolo2007dynamics}.}

\textcolor{black}{In this paper, we study how normal stress affects both advancing and receding contact line motion of partially wetting fluids. Through our analysis we demonstrate the strong effect of viscoelastic normal stress on receding contact lines which was previously observed in the experiments of~\citet{bartolo2007dynamics}. Our paper is based on an expansion of the stress to second-order in the shear rate, an approximation which exhibits a normal stress effect. The so-called second-order fluid is appropriate for modelling slow and steady flows in the weak viscoelasticity limit~\citep{tanner2000engineering, morozov2015introduction, decorato16b}. This constitutive relation} was recently given a long-wave expansion \citep{datt2022thin}, yielding a thin film equation that can be used to describe contact line motion in the presence of normal stress. The approach \textcolor{black}{resembles somewhat} the analysis by \citet{han2014theoretical}, who used corner solutions of the second-order fluid to estimate how normal stress affects the pressure inside the liquid. However, the thin film description by \citet{datt2022thin} results from a systematic long-wave expansion (naturally involving the gradient of pressure, rather than the pressure), and has the additional benefit \textcolor{black}{of applying} beyond the no-slip condition to relieve the moving contact line singularity. We focus on the case of partial wetting using a Navier-slip condition as a microscopic regularisation, so that the slip length $\lambda_s$ appears as the inner scale. The central result of \textcolor{black}{this} paper consists of the modified form of the Cox-Voinov law,

\begin{equation}
	\label{eq:intro_modified_cox}
	\theta_{\mathrm{app},o}^3 = \theta_e^3 - \frac{3}{4} \frac{\psi U^2\theta_e}{\gamma \lambda_s} \pm 9 \mathrm{Ca} \ln{\left( \frac{\ell_o}{\lambda_s}\right)},
\end{equation}
which now exhibits an explicit dependence on the normal stress coefficient $\psi$. A dependence of the Cox-Voinov relation on $\sim U^2$ was also suggested by \citet{Kim2015Mass}, but in that study the sign was proposed to be different for advancing (+) and receding (-) motion. Here we find that normal stress always leads to a smaller macroscopic contact angle. Comparing the first two terms on the right hand side of \eqref{eq:intro_modified_cox}, we identify a dimensionless normal stress parameter, 
\begin{equation}
	\label{eq:Ndef}
	N = \frac{\psi U^2}{\gamma \lambda_s \theta_e^2}.
\end{equation}
This dimensionless number $N \sim (U/U_\psi)^2$ can be seen as the viscoelastic equivalent of the capillary number, comparing the contact line speed to the intrinsic velocity $U_\psi$ defined in \eqref{eq:upsi}. Indeed, as we will show, the derivation of \eqref{eq:intro_modified_cox} requires $N \ll 1$ in a way similar to ${\rm Ca} \ll 1$. The physical picture emerging from the viscoelastic Cox-Voinov theory is \textcolor{black}{sketched} in Figure~\ref{fig:fig1}(c). The modification with respect to the Newtonian case appears only at small scales, where the bending of the interface due to normal stress dominates over the viscous bending. The form \textcolor{black}{of the contact angle law}~\eqref{eq:intro_modified_cox} \textcolor{black}{can then be} interpreted in terms of an apparent microscopic angle $\theta_{\mathrm{app},i}$, as shown in figure~\ref{fig:fig1}(c), of the form
\begin{equation}
	\theta_{\mathrm{app},i}^3 = \theta_e^3 \left(1 - \frac{3}{4} N \right).
\end{equation} 
This angle decreases with viscoelasticity, both for advancing and receding contact lines, as the viscoelastic normal stress parameter depends on $U^2$. We recall \textcolor{black}{that} \eqref{eq:intro_modified_cox} is derived under the assumption of weak viscoelasticity, i.e. $N \ll 1$. Importantly, however, when $N \gtrsim 1$, another regime of strong viscoelasticity can emerge. As a result, the complete wetting solution of \eqref{eq:Boudaoud} \citep{boudaoud2007non} is recovered, even when the substrate is partially wetting. Section~\ref{sec:modified_cox_voinov} provides the detailed derivations of these results. Subsequently, Section~\ref{sec:applications} explores the consequences of viscoelasticity for drop spreading, drop retraction and dip-coating. Finally, the paper closes with a discussion in Section~\ref{sec:discussion}.


\section{Modified Cox-Voinov law}\label{sec:modified_cox_voinov}
\subsection{Second-order fluid thin-film equation}
We investigate a moving three-phase contact line, by considering a liquid film on a rigid surface. We choose a reference frame in which the contact line is at a fixed position $x=0$ and the solid surface is moving at a speed $U$. We treat the problem with the lubrication approximation, which assumes relatively small angles. To include the effect of normal stress differences, we use the simplest viscoelastic fluid model containing normal stresses, \textit{i.e.} the second-order fluid model \citep{tanner2000engineering}. This model offers a good description of polymer solution rheology for steady flows in the weakly viscoelastic limit. 

The lubrication equation for steady contact line motion with Navier-slip boundary condition has been derived recently for the second-order fluid \citep{datt2022thin}. Denoting the interface profile as $h(x)$ (see Figure~\ref{fig:fig1}), the lubrication equation reads
\begin{equation}
	\label{eq:dimensional_lubrication_eqn_without_g}
	\gamma h''' = -\frac{3 \eta U}{h \left(h +3 \lambda_s \right)} - \frac{3}{4} \left( \psi \left(\frac{U}{h +3 \lambda_s } \right)^2 \right)' ,
\end{equation}
where $\lambda_s$ is the Navier-slip length. The introduction of a molecular mechanism, here \textcolor{black}{the} slip length, is necessary to regularise the well-known moving contact line singularity \citep{huh1971hydrodynamic,bonn2009wetting,snoeijer2013moving}. Setting $\psi=0$, we recover the standard Newtonian lubrication equation \citep{duffy1997third, oron1997long}. The extra viscoelastic term in equation~\eqref{eq:dimensional_lubrication_eqn_without_g} is of the form of the gradient of normal stress, which is quadratic in shear rate. A very similar lubrication equation was \textcolor{black}{obtained previously} in \citet{boudaoud2007non}, using scaling estimates for the components of the stress tensor. The \textcolor{black}{resulting} equation differs from \eqref{eq:dimensional_lubrication_eqn_without_g} in two ways: slip is ignored and the prefactor in front of the viscoelastic term is not correct. We emphasise here that the second-order fluid thin-film equation is asymptotically equivalent to the Oldroyd-B model for weakly viscoelastic steady flows (see Appendix in \citet{datt2022thin}), ensuring the validity of \eqref{eq:dimensional_lubrication_eqn_without_g}. Importantly, the viscoelastic term scales as $1/h^3$ whereas the viscous term scales as $1/h^2$. Hence, we anticipate the normal stress to be negligible in the large $h$ limit with respect to the viscous term. The viscoelastic term is subdominant in the region far from the contact line, where viscous shear and capillary forces balance (see region (c$_{ii}$) in figure~\ref{fig:fig1}). A consequence is that the structure of the Cox-Voinov law still holds with normal stress, and normal stress effects are important only at small scales.

We thus focus on the inner region near the contact line, in which the typical film thickness scale is $\lambda_s$. We therefore rescale $h$ and $x$ as 
\begin{equation}
	X = \frac{x \theta_e}{3 \lambda_s }, \quad \quad H(X) = \frac{h}{3 \lambda_s}, \quad \quad \lambda = \frac{\lambda_s}{\theta_e}, \quad \quad \delta = \frac{3 \eta U}{\gamma \theta_e^3} = \frac{3 \mathrm{Ca}}{\theta_e^3},
\end{equation}
where the lateral scale has been rescaled by $3\lambda_s/\theta_e$ to enforce slopes of order 1 and to scale out the equilibrium angle. We also introduced a rescaled slip length $\lambda$ and capillary number $\delta$ following \citet{eggers2005existence}. The lubrication equation then becomes
\begin{equation}
	\label{eq:Hppp_non_dimensional}
	H''' = -\delta \left( \frac{1}{H^2 +H} - \frac{\ell_\mathrm{VE}}{3 \lambda} \frac{ H'}{\left( H+1\right)^3} \right), 
\end{equation}
where viscoelastic effects are quantified by the dimensionless ratio $\ell_\mathrm{VE}/\lambda$. 

Equation~\eqref{eq:Hppp_non_dimensional} holds both for advancing ($\delta>0$) and receeding ($\delta<0$) contact lines. In this section, we focus on advancing contact lines, for which the boundary conditions are \citep{eggers2005existence}:
\begin{equation}
	\label{eq:boundary_conditions}
	H(0)=0, \quad \quad H'(0)=1, \quad \quad H''(\infty) =0,
\end{equation}
fixing the position of the contact line and the equilibrium angle condition, respectively. The third condition $H''(\infty)=0$, which can only be imposed for advancing contact lines \citep{duffy1997third}, matches any outer solution that has an asymptotically small curvature toward the contact line. 

\begin{figure}
	\centerline{\includegraphics{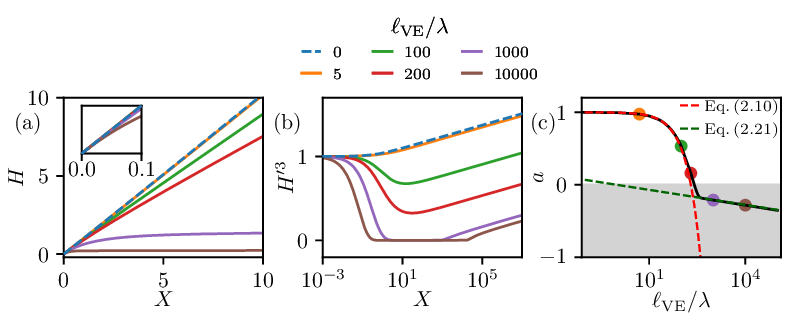}}
	\caption{\textcolor{black}{Advancing contact line:} (a) Interface profile $H(X)$ obtained from numerical integration of \eqref{eq:Hppp_non_dimensional},\eqref{eq:boundary_conditions}, with reduced capillary number  $\delta = 10^{-2}$. \textcolor{black}{The} inset shows a zoom near the contact line, \textcolor{black}{illustrating} that all profiles \textcolor{black}{have} the same microscopic contact angle. (b) Cube of the interface slope $H'(X)$, as function of distance from the contact line. Different colors correspond to varying normal stress effect quantified by $\ell_\mathrm{VE}/\lambda$. (c) The parameter $a$, obtained from fitting Equation~\eqref{eq:H_prime_3} to the large-$X$ limit of $H'(X)$ is plotted as a function of $\ell_\mathrm{VE}/\lambda$. The dots indicate the cases of figures (a) and (b) using the same color code. For $a>0$, the parameter $a=(\theta_{\mathrm{app},i}/\theta_e)^3$ can be interpreted in terms of the apparent microscopic contact angle, and is well described by \eqref{eq:a} (red dashed line). A regime of ``apparent complete wetting" emerges when $a<0$, captured by \eqref{eq:a_apparent_complete_wetting} (green dashed line).}
	\label{fig:fig2}
\end{figure}	 

We perform direct numerical integration of the system \eqref{eq:Hppp_non_dimensional}-\eqref{eq:boundary_conditions}, by using the continuation code \textit{AUTO-07P} \citep{doedel2007auto}. Typical profiles $H(X)$ are shown in Figure~\ref{fig:fig2}(a) for $\delta = 10^{-2}$ and changing the ratio $\ell_\mathrm{VE}/\lambda$ to vary the strength of normal stresses. The normal stress effects tend to lower the slope of the interface profile, though each of the profiles exhibits the same microscopic contact angle $H'(0)=1$ (see Figure~\ref{fig:fig2}(b)). Asymptotically, in the large-$X$ limit, the interface slopes cubed all follow the same behavior, as
\begin{equation}
	\label{eq:H_prime_3}
	H^{\prime 3} \simeq a + 3 \delta \ln{\left( e X \right)},
\end{equation}
and differ by a vertical offset position denoted $a$, which depends on $\ell_\mathrm{VE}/\lambda$, as shown in figure~\ref{fig:fig2}(c). This asymptotic behavior relates to the Voinov solution, which is classically found in moving contact line problems, reflecting the visco-capillary balance (see region (c$_{ii}$) in figure~\ref{fig:fig1}). We introduce in \eqref{eq:H_prime_3} a factor $e= \exp({1})$ inside the logarithm for convenience, so that the Newtonian solution corresponds to $a=1$ (as shown below). Hence, as anticipated, the normal stress effect does not modify the large scale asymptotic behavior of the inner solution, but induces a non-trivial offset in the Voinov solution. In the next two subsections, we determine the offset $a$ by using asymptotic expansions, in the weak and strong viscoelastic limit respectively.


\subsection{Weak viscoelasticity: apparent microscopic angle} \label{sec:apparent_microscopic_angle}
The standard approach to determine the interface profile near the contact line analytically is via a perturbation expansion of \eqref{eq:Hppp_non_dimensional} in the small parameter $\delta$ \citep{hocking1983spreading, eggers2005existence}: 
\begin{equation}
	\label{eq:perturbation_expansion}
	H(X) = H_0(X) + \delta H_1(X) + O(\delta^2).
\end{equation}
Here, supposing weak viscoelasticity, we \textcolor{black}{assume explicitly} the ratio $\ell_\mathrm{VE}/\lambda$ to be of order $\sim 1$. The zeroth-order solution satisfies $H_0'''=0$, which from the boundary conditions \eqref{eq:boundary_conditions} gives $H_0 = X$. At the first order in $\delta$, the interface profile follows
\begin{equation}
	\label{eq:H_1ppp_non_dimensional}
	H_1''' = - \frac{1}{X^2 + X} + \frac{\ell_\mathrm{VE}}{3 \lambda} \frac{1}{\left( 1 + X \right)^3}, 
\end{equation}
with boundary conditions $H_1'(0)=0$ and $H_1''(\infty)=0$. Integration of equation~\eqref{eq:H_1ppp_non_dimensional} then gives
\begin{equation}
	H_1' = - X \ln{X} + (1+X) \ln{(1+X)} + \frac{\ell_\mathrm{VE}}{6 \lambda} \left( \frac{1}{1+X} - 1\right) \simeq  \ln{X} + 1 - \frac{\ell_\mathrm{VE}}{6 \lambda} ,
\end{equation}
where the last step represents the $X \to \infty$ asymptotic behavior. Combined with the zeroth-order solution, the large-$X$ asymptotic behavior of the interface slope in the small-$\delta$ limit is
\begin{equation}
	\label{eq:Hp_pertub_delta}
	H' \simeq 1 + \delta \left( \ln{X} + 1 - \frac{\ell_\mathrm{VE}}{6 \lambda} \right) + O(\delta^2).
\end{equation}
Raising this result to the third power and comparing to equation~\eqref{eq:H_prime_3}, we obtain the sought-after offset constant
\begin{equation}
	\label{eq:a}
	a = 1 - \frac{1}{2} \frac{\delta \ell_\mathrm{VE}}{\lambda}.
\end{equation}
Compared to the Newtonian case where $a=1$, the interface slope indeed decreases due to the presence of normal stress effects. The dimensional form of equation~\eqref{eq:H_prime_3} reads
\begin{equation}
	\label{eq:h_prime_3_in_terms_of_Ca}
	h^{\prime 3} = \left(\theta_{\mathrm{app},i} \right)^3 + 9 \mathrm{Ca} \ln{\left( \frac{e x \theta_e}{3 \lambda_s} \right)}.
\end{equation}
Here we introduce the apparent microscopic angle $\theta_{\mathrm{app},i}$ (see figure~\ref{fig:fig1}), that depends on the normal stress parameter $N$ as, 
\begin{equation}
	\label{eq:theta_app_i}
	\left(\theta_{\mathrm{app},i} \right)^3= a \theta_e^3 = \theta_e^3 \left( 1 - \frac{3}{4} N \right),
\end{equation}
where we used $\ell_\mathrm{VE}/\lambda = 3 N/\left( 2 \delta \right)$. We identify equation~\eqref{eq:h_prime_3_in_terms_of_Ca} as a modified Cox-Voinov solution where the equilibrium Young contact angle $\theta_e$ has been modified to an apparent microscopic angle $\theta_{\mathrm{app},i}$ (see \eqref{eq:theta_app_i}). This result corresponds to equation~\eqref{eq:intro_modified_cox} in Section~\ref{sec:introduction}. 

The numerical data agree very well with this prediction for small to intermediate values of $\ell_\mathrm{VE}/\lambda$ (see red dashed lines in Figure~\ref{fig:fig2}(c)). However, a significant deviation is observed at large $\ell_\mathrm{VE}/\lambda$, which coincides with $a < 0$ (grey zone in Figure~\ref{fig:fig2}(c)). To understand the breakdown of \eqref{eq:a}, we recall that the underlying perturbation expansion assumed both $H'$ and $\ell_\mathrm{VE}/\lambda$ to remain of order unity, which is clearly violated at large $\ell_\mathrm{VE}/\lambda$. Instead, one observes the formation of a flat film region where $H' \simeq 0$ in the large viscoelastic limit (see Figure~\ref{fig:fig2}(a) \& (b)), indicating onset of a new regime. The transition between the two regimes occurs when the apparent microscopic angle goes to zero $\theta_{\mathrm{app},i} \to 0$, which occurs when
\begin{equation}
\label{eq:transition_advancing}
	\theta_e^3 \sim \frac{\psi U^2\theta_e}{\gamma \lambda_s}  \implies \frac{\ell_\mathrm{VE}}{\lambda} \sim \frac{1}{\delta}.
\end{equation}
This estimate is consistent with Figure~\ref{fig:fig2}(c), where $\delta = 10^{-2}$, for which the perturbation expansion~\eqref{eq:a} breaks down around $\ell_\mathrm{VE}/\lambda \sim 1/\delta \sim 10^2$. The next section is dedicated to a description of the new regime at large $\ell_\mathrm{VE}/\lambda$, named \textit{strong viscoelasticity}.

\subsection{Strong viscoelasticity: (apparent) complete wetting} \label{sec:apparent_complete_wetting}
We interpret the behavior at large $\ell_\mathrm{VE}/\lambda$ as a regime of ``apparent complete wetting". Indeed, the purple and brown solutions in Figure~\ref{fig:fig2}(b) exhibit an extended flat zone with $H' \approx 0$, resembling a situation of complete wetting condition, even though $H'(0)=1$. In the flat film region, the film curvature is very small, such that the capillary term is subdominant. Hence, the governing balance of \eqref{eq:dimensional_lubrication_eqn_without_g} involves the viscous and viscoelastic term, from which one identifies $\ell_\mathrm{VE}$ as the natural lateral length scale. Physically, this implies that the information of the partial wetting nature of the problem, encoded in $\theta_e$, is lost across the flat film: as it only modifies the profiles at a length scale $\sim \lambda$ (see Figure~\ref{fig:fig2}(b)), but this is ‘‘screened'' by the flat wetting film on a length scale $\ell_\mathrm{VE}$.

We proceed in the asymptotic analysis using $\ell_\mathrm{VE}$ as the relevant length on the flat film region. We assume, and validate \textit{a posteriori}, that we can drop the slip length $\lambda_s$ from \eqref{eq:dimensional_lubrication_eqn_without_g}, in the large $\ell_\mathrm{VE}/\lambda$ limit. Hence, we introduce the rescaling 
\begin{equation}
	\xi = \frac{x}{\ell_\mathrm{VE}}, \quad \quad \mathcal{H} = \frac{h}{\ell_\mathrm{VE} (3\mathrm{Ca})^{1/3}},
\end{equation}
for which equation~\eqref{eq:dimensional_lubrication_eqn_without_g} becomes
\begin{equation}
	\label{eq:mathcalHppp_without_slip_eqn}
	\mathcal{H}''' = -\frac{1}{\mathcal{H}^2} + \frac{\mathcal{H}'}{\mathcal{H}^3}.
\end{equation}
Here we dropped the slip terms, which is consistent as long as $\ell_\mathrm{VE}/\lambda \gg \delta^{-1/3}$, which is indeed satisfied in the strong viscoelastic limit. We notice that equation~\eqref{eq:mathcalHppp_without_slip_eqn} resembles the thin-film equation with a disjoining pressure (see page 398 in \citet{eggers2015singularities}), which is used to model contact lines in complete wetting. Within this analogy, the effective ``disjoining pressure'' is $-3\psi U^2 / (4h^2)$, similar to the standard van der Waals disjoining pressure $-A/(6\pi h^3)$, where $A$ is the Hamaker constant. We employ boundary conditions
\begin{equation}
	\label{eq:boundary_conditions_new_scaling}
	\mathcal{H}(-\infty)=0, \quad \quad \mathcal{H}'(-\infty)=0, \quad \quad \mathcal{H}''(+\infty) =0,
\end{equation}
where the first two conditions correspond to the formation of the film, replacing the partial wetting condition at $\xi=0$. Similarly to the Section~\ref{sec:apparent_microscopic_angle}, since viscoelasticity only acts at small scales, the large distance asymptotic behavior is given by the Voinov solution (see \eqref{eq:H_prime_3}), which using the new scales reads
\begin{equation}
	\mathcal{H}'^3 \simeq 3 \ln \left(\frac{\xi}{\xi_0} \right).
\end{equation}
The offset $a$ has been replaced by an unknown constant $\xi_0$, which remains to be found. 

The system \eqref{eq:mathcalHppp_without_slip_eqn}-\eqref{eq:boundary_conditions_new_scaling} is exactly the problem analysed in \citet{boudaoud2007non} up to a trivial rescaling, which we now understand to emerge as the limit of large $\ell_\mathrm{VE}/\lambda$. In the flat film region, the curvature is nearly zero and the capillary term drops out, such that the dominant balance involves viscous stress and normal stress. Towards the flat film, equation~\eqref{eq:mathcalHppp_without_slip_eqn} reduces to
\begin{equation}
	\mathcal{H}' \simeq \mathcal{H},
\end{equation} 
to leading order. The viscous-viscoelastic solution takes the form of an exponentially growing solution $\mathcal H \simeq  e^{\xi}$, with \textcolor{black}{a} prefactor that can be set to unity, owing to the translational invariance of \eqref{eq:mathcalHppp_without_slip_eqn} and \eqref{eq:boundary_conditions_new_scaling}. A more refined analysis of the asymptotic behavior of $\mathcal{H}$ in the $\xi \to -\infty$ limit leads to the asymptotic expression \citep{boudaoud2007non}: 
\begin{equation}
	\label{eq:initial_condition_apparent_complete_wetting}
	\mathcal{H} \simeq e^{\xi} + c_{+} e^{ \frac{2}{3} e^{-3\xi/2}} + c_{-} e^{ - \frac{2}{3} e^{-3\xi/2}},
\end{equation}
where $c_\pm$ are constants; using the boundary condition for $\xi \to -\infty$ leads to $c_{+} =0$. We solve equation~\eqref{eq:mathcalHppp_without_slip_eqn} numerically, using equation~\eqref{eq:initial_condition_apparent_complete_wetting} as an initial condition, where $c_-$ is treated as a shooting parameter to ensure $\mathcal{H}''(\xi \to \infty) =0$. Fitting the numerically-obtained slope $\mathcal{H}'(\xi)$ in the large-$\xi$ limit, typically $\xi = 10^8$, with the Voinov solution \citep{bender1999advanced, eggers2015singularities}, we find a universal value of $\xi_0 \approx 1.16$ \textcolor{black}{(see \citet{snoeijer2010asymptotic} for details of the numerical calculation)}. 

In dimensional units, the contact line solution at large $x$ gives
\begin{equation}
	\label{eq:hppp_app_complete_wetting}
	h'^3 = 9 \mathrm{Ca} \ln{\left( \frac{x}{\ell_\mathrm{VE} \xi_0}\right)}.
\end{equation}
This has the same form of the Cox-Voinov solution for complete wetting, \textit{i.e.} when $\theta_e=0$, where $\ell_\mathrm{VE}$ now offers the regularisation length, replacing the microscopic scale. This was also demonstrated in \citep{boudaoud2007non}, but \textcolor{black}{we now} provide the prefactor. In terms of the constant $a$, as defined by \eqref{eq:H_prime_3}, we find 
\begin{equation}
	\label{eq:a_apparent_complete_wetting}
	a = 3 \delta \ln \left( \frac{3 \lambda}{e \ell_\mathrm{VE} \xi_0}  \right),
\end{equation}
which agrees well with the numerical solution at large $\ell_\mathrm{VE}/\lambda$ (Figure~\ref{fig:fig2}(c), green dashed line). It justifies \textit{a posteriori} that the equilibrium angle $\theta_e$ does not play a role in this regime.

\section{Applications}
\label{sec:applications}
We now apply the viscoelastic contact line theory from the previous section to two paradigmatic wetting problems, namely drop spreading and dip coating. Classically, the interface profile in these problems can be described using matched asymptotics, where the Voinov contact line solution corresponds to the inner solution, matched to an outer solution at macroscopic scale. In this section, we briefly recall the outer solutions of the two problems, which are unaltered by viscoelasticity, and discuss the matching to the newly-derived viscoelastic inner solutions. 

In applications such as drop spreading the contact line velocity $U$ is not a control parameter. Instead, the contact line velocity is determined through the matching procedure. Using the Cox-Voinov framework, an intrinsic velocity scale $U^*$ naturally arises by comparing the viscous and capillary term as 
\begin{equation}
	U^* = \frac{\gamma \theta_e^3}{\eta},
\end{equation} 
which we will use as a rescaling velocity below. In these scaled units, we will see the appearance of a new dimensionless parameter

\begin{equation}\label{eq:N0def}
	N_0 = \left( \frac{U^*}{U_\psi} \right)^2 = 
	\frac{\psi \gamma \theta_e^4}{\eta^2 \lambda_s }.
\end{equation}
This parameter governs the relative importance of viscosity and normal stress in the context of partially wetting flows. Alternatively, $N_0$ can be interpreted as a scaled normal stress coefficient, made dimensionless only with material parameters. 
\subsection{Drop spreading and retraction}\label{sec:spreading}

We study the spreading (and retraction) of a drop, of volume $\Omega$, over a solid surface (Figure~\ref{fig:fig3}). We consider drops smaller than the capillary length so that gravity can be ignored. The outer solution of the interface profile is obtained away from the contact line, where viscous and viscoelastic forces are not dominant. Therefore, at a given time, the shape adopted by the drop is close to the equilibrium shape of a static drop $h_0(r)$ as, 
\begin{equation} 
	\label{eq:static_profile_drop}
	h_0(r)= \frac{2 \Omega}{\pi r^2} \left[ 1- \left( \frac{r}{R}\right)^2 \right],
\end{equation}
where the drop meets the solid at the apparent angle $\theta_{\mathrm{app},o} = -h_0'(r=R)$, which reads
\begin{equation}
\label{eq:theta_app_drop}
	\theta_{\mathrm{app},o}= \frac{4 \Omega}{ \pi R^3}.
\end{equation}
As discussed before, close to the contact line (to be precise, the outer asymptotics of the inner region), the slope of an advancing contact line solution exhibits a logarithmic dependence with the distance. Hence the static equilibrium profile \eqref{eq:static_profile_drop} cannot be matched directly to the Cox-Voinov theory. To resolve this issue, \citet{hocking1983spreading} showed that the first order correction to the outer solution in capillary number, defined as $\eta \dot{R} /\gamma$ where $\dot{R}$ is the radius rate of change, takes the asymptotic form
\begin{equation}
	\label{eq:hppp_drop_spreading_outer}
	h'^3(x)= \theta_{\mathrm{app},o}^3 + 9 \frac{\eta \dot{R}}{\gamma}\ln{\left( \frac{x}{R b}\right)},
\end{equation}
for $x=R-r \ll R$ (see figure~\ref{fig:fig3}(a)), and where $b=1/\left(2e^2 \right)$. This solution is to be matched to the newly-derived viscoelastic inner solutions\textcolor{black}{~\eqref{eq:h_prime_3_in_terms_of_Ca} and~\eqref{eq:hppp_app_complete_wetting}.} 

\subsubsection{Spreading with weak viscoelasticity: partial wetting}
\label{subsec:partialwetting}
\begin{figure}
	\centerline{\includegraphics{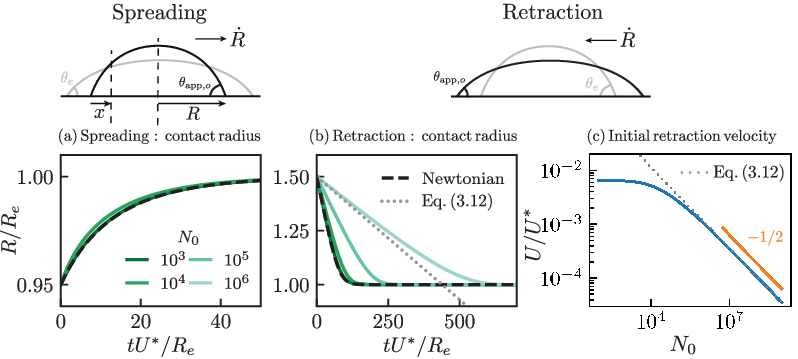}}
	\caption{Drop spreading (a) and retraction (b)\textcolor{black}{-(c)}. The schematic indicates the case, where the initial drop is drawn in black and the equilibrium shape in gray. The drop \textcolor{black}{contact} radius rescaled by the equilibrium \textcolor{black}{contact} radius is plotted as a function of dimensionless time for various $N_0$, corresponding to numerical solutions of \eqref{eq:drop_spreading_ODE_integrated}, where $c_1 = 10^7$. The black dashed lines show the Newtonian solution for $N_0=0$, and the gray dotted line displays the viscoelastic retraction velocity, corresponding to \eqref{eq:drop-retraction_viscoelastic} evaluated at the initial time. \textcolor{black}{(c) The initial retraction velocity scaled by $U^*$ as a function of $N_0$, computed from~\eqref{eq:drop_spreading_ODE_dimensionless} using $R_{\textit{initial}}/R_e=1.5$ (as in panel (b)). The orange line in large $N_0$ regime corresponds to $U \sim \psi^{-1/2}$ scaling as expected from equation~\eqref{eq:drop-retraction_viscoelastic}.}}
	\label{fig:fig3}
\end{figure}
For the partial wetting case, the drop spreads until the contact angle reaches the equilibrium contact angle $\theta_e$, or equivalently an equilibrium radius $R_{e}= \left( 4 \Omega/ \left(\pi \theta_e \right) \right)^{1/3}$. For small capillary number, we can match the logarithmic dependencies of the outer solution~\eqref{eq:hppp_drop_spreading_outer} to the modified Cox-Voinov solution~\eqref{eq:h_prime_3_in_terms_of_Ca}, which gives
\begin{equation}
	\left(\theta_{\mathrm{app},o} \right)^3 = \left(\theta_{\mathrm{app},i} \right)^3 + 9 \frac{\eta \dot{R}}{\gamma} \ln{\left( \frac{R \theta_{e}}{6 e \lambda_s} \right)}.
\end{equation}
As usual, the matching procedure connects the microscopic angle to the macroscopic angle, but now the microscopic angle contains viscoelastic effects. Inserting $\theta_{\mathrm{app},o}$ from the static solution~\eqref{eq:theta_app_drop} and $\theta_{\mathrm{app},i}$ from \eqref{eq:theta_app_i}, we obtain an ODE for the drop radius $R(t)$, as 
\begin{equation}
	\label{eq:drop_spreading_ODE}
	\left(\theta_{\mathrm{app},o} \right)^3 = \left(\frac{4 \Omega}{ \pi R^3} \right)^3 = \theta_e^3 - \frac{3}{4}\frac{\psi \dot{R}^2 \theta_e}{\gamma \lambda_s} + 9 \frac{\eta \dot{R}}{\gamma} \ln{\left( \frac{R \theta_{e}}{6 e \lambda_s} \right)}.
\end{equation}
Finally, using dimensionless variables with bar as $\bar{R} = R/R_e$ and $\bar{t} = t / (R_e/U^*)$, we can rewrite \eqref{eq:drop_spreading_ODE} as 
\begin{equation}
	\label{eq:drop_spreading_ODE_dimensionless}
\frac{3}{4} N_0 \dot{\bar{R}}^2 - 9 \ln{\left( c_1\bar{R}  \right)} \dot{\bar{R}} + \left( \frac{1}{\bar{R}^9} -  1  \right) = 0,
\end{equation}
with $c_1 = (R_e \theta_e)/(6 e \lambda_s)$. Here we recover the explicit dependency on $N_0$, which is the material parameter that for a given fluid quantifies normal stress effect. 

Equation~\eqref{eq:drop_spreading_ODE_dimensionless} is a second-order polynomial equation for $\dot{\bar{R}}$, which we first solve to isolate $\dot{\bar{R}}$, leading to 
\begin{equation}
\label{eq:drop_spreading_ODE_integrated}
	\dot{\bar{R}} = \frac{6\ln\left( c_1 \bar{R}  \right)}{N_0} \left[1 - \sqrt{1 - \frac{N_0}{27 \ln^2\left( c_1 \bar{R}  \right)}\left(\frac{1}{\bar{R}^9}-1\right)}\right],
\end{equation}
\textcolor{black}{and then integrate numerically}. We notice that the second-order polynomial equation does not have a solution in real space for negative discriminant, which imposes a condition $27\ln^2\left( c_1\bar{R}  \right) >  N_0 (1/\bar{R}^9-1)$. Hence, in the spreading case, \textit{i.e.} when $\bar{R}(\bar{t}=0)<1$, we cannot find solutions for arbitrary large values of $N_0$. Mathematically speaking, the reason is that the apparent macroscopic angle versus $U$ has a maximum value when using the modified Cox-Voinov law. Therefore, if the outer geometry imposes an apparent macroscopic angle that is larger than this maximum, no contact-line speed can be found. The lack of solution is an artefact of the weak viscoelastic solution, which \textcolor{black}{as also mentioned in the Discussion} is resolved by admitting the strong viscoelastic solution \eqref{eq:hppp_app_complete_wetting}. 

In figure~\ref{fig:fig3}(a) we show a typical solution of \eqref{eq:drop_spreading_ODE_integrated}, with $\bar{R}(\bar{t}=0) = 0.95$ as an initial condition. Qualitatively, the normal stress effects speed up the spreading dynamics, although the quantitative change is very weak. This corroborates the conclusions of drop impact with polymer solutions, showing a weak dependence on the addition of polymers~\citep{bartolo2007dynamics,gorin2022universal,sen2022elastocapillary}.

\subsubsection{Spreading with strong viscoelasticity: (apparent) complete wetting}
As shown in section~\ref{sec:apparent_complete_wetting}, the advancing contact line dynamics at large velocities deviates from the modified Cox-Voinov law used in the previous subsection. In the drop spreading problem, if the initial radius is small, or equivalently the apparent angle is much larger than the equilibrium angle, then the contact-line speed is large and lies within the strongly viscoelastic regime. In this case, the inner solution for partially wetting substrates becomes completely independent of the equilibrium angle, and takes the form \eqref{eq:hppp_app_complete_wetting}. Repeating the matching procedure with this strongly viscoelastic inner solution, the drop radius  follows from
\begin{equation}
	\label{eq:theta_app_global_problem_matching}
	\theta_{\mathrm{app},o}^3 = 9 \frac{\eta \dot{R}}{\gamma} \ln{\left( \frac{R b}{\ell_\mathrm{VE} \xi_0}\right)} = 9 \frac{\eta \dot{R}}{\gamma} \ln{\left( \frac{2\eta R b}{\psi \dot{R} \xi_0}\right)}.
\end{equation}
This result is identical to that of a completely wetting viscoelastic drop\textcolor{black}{; and it is} of the same form as derived in \citet{boudaoud2007non}.  

During the spreading, which we initially assumed to be rapid, the contact line velocity will decrease in time. For partially wetting drops, this means that upon approaching the equilibrium angle, the strongly viscoelastic regime will give way to the weak regime discussed in the preceding paragraph. 

For completetly wetting surfaces, however, the drop radius follows a scaling law $R\propto t^{1/10}$ with logarithmic corrections\textcolor{black}{, which is analogous to the Newtonian dynamics known as Tanner's law \citep{bonn2009wetting, eggers2015singularities}}. Using $R \propto t^{1/10}$ as an Ansatz, the ratio between contact line speed and drop radius inside the logarithm in \eqref{eq:theta_app_global_problem_matching} can be approximated as $\dot{R}/R \approx 1/(10t)$. Inserting \eqref{eq:theta_app_drop} into~\eqref{eq:theta_app_global_problem_matching}, and integrating, we obtain a modified Tanner's law as
\begin{equation}
	\label{eq:viscoelastic-Tanner}
	R \approx \left( \frac{10}{9} \left( \frac{4 \Omega}{ \pi} \right)^3 \frac{\gamma}{\eta \ln{\left( c_2\frac{\eta t}{\psi}\right)}}  t \right)^{1/10}.
\end{equation}
where $c_2= 10/(e^2\xi_0)$. We note that the same equation has been derived using scaling arguments in \citet{rafai2004spreading,boudaoud2007non}, but we now provide a prefactor as derived for the second-order fluid. Again, the spreading of drops is accelerated by normal stress effects with respect to the Newtonian dynamics via the logarithmic term in \eqref{eq:viscoelastic-Tanner}. 

\subsubsection{Drop retraction}
If the initial drop radius is above the equilibrium radius, the drop is retracting (see figure~\ref{fig:fig3}(b)), for instance after an impact on a surface. Even though the contact line is now receding, the same framework as Section~\ref{subsec:partialwetting} can be used as long as the change in contact angle is sufficiently small (cf. \cite{eggers2005existence} and the discussion on dip-coating). By consequence, \eqref{eq:drop_spreading_ODE_dimensionless} also governs the retraction dynamics of a partially wetting drop, but $\dot{\bar R}$ is now negative. In the retraction case, the discriminant of the second-order polynomial equation is always positive such that one can find a solution for any $N_0$. Interestingly, viscoelastic effects now oppose the motion and slow down the retraction (see figure~\ref{fig:fig3}(b)). In the large viscoelastic limit, for $N_0 \to \infty$, the viscous stress becomes negligible and the retraction velocity is set by balancing the normal stress and the capillary term of \eqref{eq:drop_spreading_ODE}, which gives in dimensional form
\begin{equation}
\label{eq:drop-retraction_viscoelastic}
\dot{R}^2 = \frac{4 \gamma (\theta_e^3-\theta_{\mathrm{app},o}^3)\lambda_s}{3\psi \theta_e}. 
\end{equation}
The dotted line in figure~\ref{fig:fig3}(b) indicates the retraction as estimated using the viscoelastic receding speed \eqref{eq:drop-retraction_viscoelastic}, with the apparent angle taken from the initial condition. This provides a very good description of the retraction velocity for large $N_0$ (the contact line only slowing down upon approaching the equilibrium radius). \textcolor{black}{The initial retraction velocity is plotted as a function of $N_0$ in figure~\ref{fig:fig3}(c). In the small viscoelastic limit, quantified by small $N_0$, the viscoelastic term of equation~\eqref{eq:drop_spreading_ODE_dimensionless} is subdominant and the retraction velocity typically scales with $U^*$. Conversly, for large viscoelasticity, the retraction velocity is set by $U_\psi \sim 1/\sqrt{\psi}$, leading to the scaling $U/U^* \sim N_0^{-1/2}$, observed at large $N_0$ in figure~\ref{fig:fig3}(c). Using the values of relevant parameters from the experiments by~\citet{bartolo2007dynamics}, we estimate $N_0$ to be in the range $\sim 10^3$ to $10^6$. The $N_0$ values chosen in figures~\ref{fig:fig3}(a)-(b) fall within this range.}

In the large viscoelastic limit, this retraction velocity scales with the inverse square root normal stress coefficient, as observed experimentally in~\citet{bartolo2007dynamics}. These authors rationalised their experimental findings by using a phenomenological normal stress thin-film model and derived a retraction velocity of similar form as the one discussed here, namely $\dot{R}^2 = \gamma (\theta_e^2 - \theta_{\mathrm{app},o}^2) \ell /(8 \psi)$, where $\ell$ is a microscopic cut-off length. Up to a prefactor, \textcolor{black}{in the limit of $\theta_{\mathrm{app},o} \to 0$,} this phenomenological equation agrees with the rigorous calculation provided here. 


\subsection{Forced wetting transition}
\begin{figure}
	\centerline{\includegraphics[width= \linewidth]{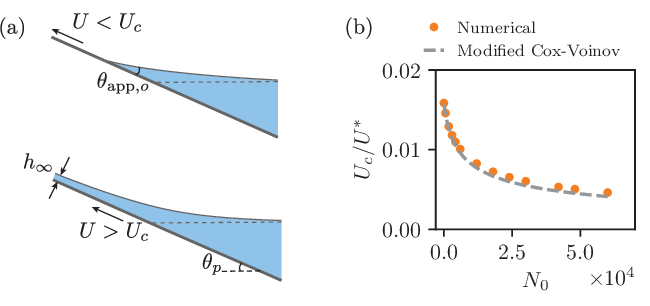}}
	\caption{Forced wetting transition: (a) Schematic of the dip-coating problem. Below the critical speed $U_c$ (top), the contact line is displaced and the apparent contact angle is $\theta_{\mathrm{app},o}$. Above the critical speed, a Landau-Levich film of thickness $h_\infty$ is entrained. (b) Critical dimensionless speed as function of dimensionless normal stress coefficient. The dots represents full scale numerical integration of \eqref{eq:lubrication-dip-coating}, with $\theta_p/\theta_e = 1$ and $\lambda_s/(\ell_\gamma\theta_e) = 10^{-4}$. The dashed line shows the asymptotic matching solution \eqref{eq:viscoelastic_wetting_transition} based on the modified Cox-Voinov law.}
	\label{fig:fig4}
\end{figure}

We now turn to dip-coating, sketched in figure~\ref{fig:fig4}(a), in which a plate is pulled out of a liquid bath at a constant speed $U$. At small speed, the liquid-air interface is modified and the position of the contact line is shifted upward with respect to the equilibrium position. Above a critical speed $U_c$, however, there is no static contact line solution anymore and a Landau-Levich liquid film is entrained. This transition is usually called forced dewetting~\citep{eggers2005existence,snoeijer2007relaxation,chan2012theory,galvagno2014continuous}. 

The question we wish to address is how $U_c$ is affected by viscoelasticity. We restrict ourselves to the case of small plate angle $\theta_p$, so that lubrication theory can be described for both the contact line and the bath. Assuming a stationary state, the interface position can then be computed from the lubrication equation as 
\begin{equation}
	\label{eq:lubrication-dip-coating}
	\gamma h''' = \frac{3 \eta U}{h \left(h +3 \lambda_s \right)} - \frac{3}{4} \left(\psi\left(\frac{U}{h +3 \lambda_s } \right)^2\right)' + \rho g \left( h' - \theta_p \right) .
\end{equation}
With respect to \eqref{eq:dimensional_lubrication_eqn_without_g}, the sign of the viscous term is flipped as the contact line recedes, and a gravitational term has been added, $\rho$ and $g$ being the liquid density and gravitational acceleration respectively. The sign of the viscoelastic term is unchanged, as it is quadratic in contact line speed. The boundary conditions are
\begin{equation}\label{eq:dipcoatingbc}
	h(0)=0, \quad \quad h'(0)=\theta_e, \quad \quad h''(\infty) =0,
\end{equation}
and are identical to those in Section \ref{sec:modified_cox_voinov}.  In the Newtonian case, solutions only exist up to a critical velocity. Likewise, we have numerically determined $U_c$ for the viscoelastic case by finding the maximum velocity for which \eqref{eq:lubrication-dip-coating}, \eqref{eq:dipcoatingbc} admit a solution. A typical  numerical result is shown as the circles in Figure~\ref{fig:fig4}(b), where we plot the normalised critical speed $U_c/U^*$ as a function of the viscoelastic parameter $N_0$. We thus observe that the normal stress effect lowers the critical speed of entrainment with respect to the Newtonian case ($N_0=0$). Like for droplet retraction dynamics, any receding contact line motion is inhibited by viscoelastic effects, now leading to an earlier entrainment.

Once again, these results can be described quantitatively from the modified inner solution for receding contact lines. The outer solution is again unaffected by viscoelasticity: far from the contact line position, the interface profile is found by balancing the hydrostatic pressure with the Laplace pressure. This leads to the equation of a static meniscus, which is entirely characterised by an apparent angle $\theta_{\mathrm{app},o}$ (see figure~\ref{fig:fig4}(a)) and which corresponds to the outer solution of the full-scale problem. The inner solution close to the contact line, at scales much smaller than the capillary length $\ell_\gamma = \sqrt{\gamma/(\rho g)}$, does not involve gravity. Importantly, and in contrast to the advancing case, the general solution of the inner problem does not have a vanishing curvature at infinity. Therefore, the matching procedure is more intricate and analytical progress can be made by invoking an intermediate zone, for $\lambda \ll x\ll \ell_\gamma$, where slip and gravity are neglected, such that \eqref{eq:lubrication-dip-coating} reduces to the visco-capillary balance $\gamma h''' = 3\eta U/h^2$. The latter equation has an exact solution ~\citep{duffy1997third}, which has a Voinov asymptotic behavior $h'^3 = -9\mathrm{Ca} \ln (x/\ell)$ at small $x$ and a constant curvature $h'' \to \mathrm{cte}$ at large $x$. As shown in \citet{eggers2005existence}, this visco-capillary solution can be matched both to the outer static meniscus and to the slip region. 

Hence, to understand the normal stress effects on dip-coating, which appear only on the slip-scale, the remaining task is to derive the slip-scale viscoelastic receding solution. Modifying the steps of Section \ref{sec:apparent_microscopic_angle} along the lines of \citet{eggers2005existence}, but now including the normal stress term, we find that the large-$x$ asymptotic behavior of the viscoelastic inner solution follows
\begin{equation}
	\label{eq:cox_voinov_forced_wetting_transition}
	h^{\prime 3} = \left(\theta_{\mathrm{app},i} \right)^3 - 9 \mathrm{Ca} \ln{\left( \frac{e x \theta_e}{3 \lambda} \right)}.
\end{equation}
This result is identical to the advancing counterpart \eqref{eq:h_prime_3_in_terms_of_Ca}, with only a change of sign in the viscous term. Importantly, the only change with respect to the Newtonian case is a modification of the microscopic angle from $\theta_e$ to the apparent microscopic angle defined in~\eqref{eq:theta_app_i}. Following \citet{eggers2005existence}, we find the equation for the critical speed:
\begin{equation}
\theta_{\mathrm{app},i}^3 
= 9 \frac{\eta U_c}{\gamma} \ln\left( \frac{(\eta U_c/\gamma)^{1/3} \ell_\gamma \theta_e}{18^{1/3} \pi [\mathrm{Ai}(s_\mathrm{max})]^2 \lambda_s \theta_p} \right),
\end{equation}
where $\mathrm{Ai}(s_\mathrm{max}) = 0.53566\dots$ is the global maximum of the Airy function. The Newtonian result is recovered by replacing the left hand side by $\theta_e^3$. Using the characteristic speed $U^*$ as a scale, the dimensionless critical velocity $\bar{U}_c = U_c/U^*$ is the solution of 
\begin{equation}
	\label{eq:viscoelastic_wetting_transition}
	\bar{U}_c = \frac{1-\frac{3}{4} N_0 \bar{U}_c^2}{9 \ln \left(c_3 \bar{U}^{1/3}_c \right)},
\end{equation}
where $c_3 = 1/(18^{1/3} \pi [\mathrm{Ai}(s_\mathrm{max})]^2)\left(\ell_\gamma \theta_e^2 \right)/\left( \lambda_s \theta_p \right) \approx 0.423 \left(\ell_\gamma \theta_e^2 \right)/\left( \lambda_s \theta_p \right) $. Here again, we recover the material parameter $N_0$ that quantifies viscoelasticity, as discussed earlier. 

Figure~\ref{fig:fig4}(b) shows excellent agreement between (\ref{eq:viscoelastic_wetting_transition}), shown as the dashed line, and the numerical solution. Hence, the decrease of the entrainment velocity with increasing viscoelasticity can be understood from the decrease of the apparent inner angle $\theta_{{\rm app},i}$. In the limit of large $N_0$, one finds $\bar U_c^2 \simeq 4/(3N_0)$, which in dimensional units reads

\begin{equation}
\label{eq:dipcoating-limit}
U^2_c \simeq \frac{4 \gamma \theta_e^2\lambda_s}{3\psi}. 
\end{equation}
Note the similarity with \eqref{eq:drop-retraction_viscoelastic} for drop retraction. This result suggests the emergence of a universal receding velocity for strong normal stress effect, independently of the outer geometry.

\section{Discussion}\label{sec:discussion}

\begin{figure}
	\centerline{\includegraphics[width=1.00 \linewidth]{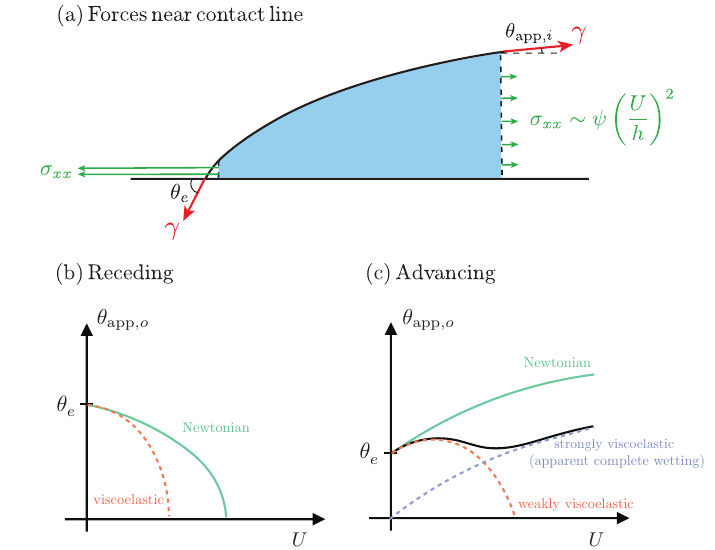}}
	\caption{\textcolor{black}{(a) Force balance on a control volume, highlighted by the blue area, in the inner region near the contact line, where capillary force $\gamma$ and the normal stress $\psi \left(U/h\right)^2$ are balanced.} Macroscopic apparent contact angle versus speed for receding (b) and advancing (c) contact lines. In the advancing case, we need to distinguish between weak and strong viscoelastic behavior. The crossover between these regimes is \textcolor{black}{indicated schematically} by the black solid line, and occurs around $U_\psi$ defined in (\ref{eq:upsi}). }
	\label{fig:summary}
\end{figure}

In this paper we have studied the effect of viscoelastic normal stresses on contact line motion. Using the lubrication equation for the second-order fluid, we have derived a modified form of the classical Cox-Voinov theory: the macroscopic apparent contact angle ($\theta_{{\rm app},o}$) is modified, due to a small-scale ``apparent inner contact angle'' ($\theta_{{\rm app},i}$) that is lowered due to viscoelasticity \textcolor{black}{both for advancing and receding contact line motion. Figure~\ref{fig:summary}(a) shows the force balance for a fluid volume in the inner region close to the contact line, with the two dominant forces: capillary force and normal stress. The contribution of the normal stress on both lateral sides can be calculated by integrating the stress across the film thickness, and scales as $\sim  \psi \left(U/h \right)^2 h= \psi U^2/h$. Hence, the left side near the contact line is dominant as $h \to 0$, and the global effect of normal stresses is to pull towards the contact line, regardless of whether the contact line is advancing or receding. This explains why the normal stress facilitates advancing and inhibits receding motions. The singularity of the viscoelastic force as $h \to 0$ is regularized by the slip length $\lambda_s$, such that the effective resulting viscoelastic force is $\psi U^2/\lambda_s$. This viscoelastic force in the inner region bends the interface and lowers the microscopic angle from $\theta_e$ to $\theta_{{\rm app},i}$. We have worked out quantitatively} the effect on drop spreading, drop retraction and dip-coating, and found fundamental differences between advancing and receding contact lines. To summarise the findings, we turn to the schematics for the macroscopic apparent angle $\theta_{{\rm app},o}$ versus speed provided in Figure~\ref{fig:summary}(b,c). 

For receding contact lines, viscosity and normal stress provide forces that act in the same direction, both opposing the contact line motion. Both viscosity and normal stress give rise to a ``bending" of the interface that tend to lower the macroscopic angle, as can be seen in Figure~\ref{fig:summary}(b). As a result, the retraction velocity of a droplet and the critical velocity for dip-coating are lowered with respect to the Newtonian case. The relative importance of viscous and viscoelastic bending is governed by the parameter $N_0$ defined in \eqref{eq:N0def}. In the case of strong viscoelasticity, the second-order fluid model predicts a universal receding velocity (\ref{eq:dipcoating-limit}), which is proportional to $U_\psi$ defined in (\ref{eq:upsi}), that gives both the critical speed for dip-coating and droplet retraction. This prediction is consistent with experiments by \citet{bartolo2007dynamics}, who in fact already introduced $U_\psi$ to explain experimental measurements of droplet retraction after impact. Specifically, they found the retraction velocity to be proportional to $1/\sqrt{\psi}$. However, they found that the microscopic length appearing in $U_\psi$ was tens of microns, which is much larger than a nanoscopic slip length. \citet{bartolo2007dynamics} remarked that the fitted lengths are comparable to the length of the fully extended polymer chains. Such large values of cutoff length may be due a shear-induced migration of the polymers which can lead to depletion layers (apparent slip lengths) in micrometers \citep{Muller1990, fang2007molecular, Graham2005, Degre2006, Shin2016}. In general, non-hydrodynamic interaction between polymers and the surface can also contribute to large slip lengths \citep{mhetar1998slip, Gupta2019}. Regardless of the molecular mechanisms at play, we expect that such effects do not alter the phenomenology discussed here, including the interpretation in terms of an apparent inner angle $\theta_{{\rm app},i} \to 0$ due to viscoelasticity. The presented theory provides a benchmark for the influence of normal stress on receding contact line motion, but molecular mechanisms \textcolor{black}{might be required to provide an effective scale of regularisation.}

For advancing contact lines, viscosity and normal stress provide forces that act in opposite directions: the normal stress actually facilitates contact line motion. We have seen that we need to distinguish between weak and strong viscoelasticity. For weak viscoelasticity, there is a non-monotonic dependency of the macroscopic angle on contact line velocity, as indicated by the orange dashed line in Figure~\ref{fig:summary}(c). The maximum arises due to a competition between viscous bending (increasing the angle) and viscoelastic bending (decreasing the angle). The crossover to the strongly viscoelastic regime is given by $U \sim U_\psi$ (see~\eqref{eq:transition_advancing}), which occurs when the normal stress effect is so strong that it bends the interface to form a precursor film of typical size $\ell_{\rm VE}$. This strongly viscoelastic regime is indicated by the dashed blue line, and can be interpreted as an ``apparent complete wetting'', since the contact line motion becomes completely insensitive to $\theta_e$. The crossover is schematically indicated by the black solid line. The exact details of the crossover from weak to strong viscoelasticity are not universal, but depend on the material parameter $N_0$ defined in \eqref{eq:N0def}. For example, the speed at which the maximum appears in the weakly viscoelastic regime scales as $U_\psi/\sqrt{N_0}$; when $N_0$ is sufficiently large the maximum will thus emerge well before it is erased by the apparent complete wetting regime.  

The presented analysis provides concrete predictions for experiments on wetting of viscoelastic liquids, rooted in a systematic long-wave expansion of the second-order fluid. Our findings suggest the possibility of a non-monotonic dependence of the contact angle with advancing velocity, which could potentially lead to instabilities. Such effects remain to be explored experimentally, but also deserve some further theoretical analysis. Namely, the calculations are based on the second-order fluid: while this lubrication model is equivalent to that obtained with Oldroyd-B fluid when viscoelasticity is introduced perturbatively \citep{Ro1995,datt2022thin}, its quantitative validity at strong viscoelasticity remains to be explored, more so when polymers in moving contact line problems have been observed to be strongly stretched \citep{smith2010effect, Shin2016}. It would therefore be of interest to study contact line motion for \textcolor{black}{a variety} of constitutive models.

\textbf{Acknowledgements.} The authors gratefully acknowledge discussions with Daniel Bonn and Satish Kumar.

\textbf{Funding.} This work was supported by NWO through VICI grant no. 680-47-632. 

\textbf{Declaration of interests.} The authors report no conflict of interest.

\textbf{Author ORCIDs.}
\\
Minkush Kansal: https://orcid.org/0000-0003-1584-3775;
\\
Vincent Bertin: https://orcid.org/0000-0002-3139-8846;
\\
Charu Datt: https://orcid.org/0000-0002-9686-1774;
\\
Jens Eggers: https://orcid.org/0000-0002-0011-5575;
\\
Jacco H. Snoeijer: https://orcid.org/0000-0001-6842-3024.

\end{document}